\documentclass[11pt]{article}
\usepackage{amssymb,amsmath,fullpage,graphicx}
\def\[{\begin{equation}}
\def\]{\end{equation}}
\def\PT{$\cal{PT}$}
\makeatletter \numberwithin{equation}{section}

\begin{document}
\title{Necessity of \PT symmetry for soliton families in one-dimensional complex
potentials}
\author{Jianke Yang \\
Department of Mathematics and Statistics \\ University of
Vermont,\\Burlington, VT 05401, USA }
\date{ }
\maketitle

\begin{abstract}
For the one-dimensional nonlinear Schr\"odinger equation with a
complex potential, it is shown that if this potential is not
parity-time (\PT) symmetric, then no continuous families of solitons
can bifurcate out from linear guided modes, even if the linear
spectrum of this potential is all real. Both localized and periodic
non-\PT-symmetric potentials are considered, and the analytical
conclusion is corroborated by explicit examples. Based on this
result, it is argued that \PT-symmetry of a one-dimensional complex
potential is a necessary condition for the existence of soliton
families.
\end{abstract}

\section{Introduction}

Nonlinear wave systems used to be divided into two main categories:
conservative systems and dissipative systems. In the former
category, the system has no energy gain or loss, and solitary waves
(or solitons in short) exist as continuous families, parameterized
by their propagation constants. A well known example is the
nonlinear Schr\"odinger equation with or without a real potential
\cite{Kivshar2003, Yang2010}. In the latter category, the system has
energy gain and loss, and solitons generally exist as isolated
solutions at certain discrete propagation-constant values, where the
energy gain and loss on the soliton are exactly balanced (such
solitons are often referred to as dissipative solitons in the
literature) \cite{Akhmediev2005}. A typical example in this latter
category is the Ginzburg-–Landau equation or short-pulse lasers (see
also \cite{Zezyulin2011}).

However, a recent discovery is that, in dissipative but parity-time
(\PT) symmetric systems, solitons can still exist as continuous
families, parameterized by their propagation constants
\cite{Musslimani2008, Wang2011, Lu2011, Abdullaev2011, He2012,
Nixon2012, Zezyulin2012a, Huang2013,Kartashov2013,Driben2011,
Alexeeva2012,Moreira2012,Konotop2012,Barashenkov2013,Kevrekidis2013,Li2011,
Zezyulin2012b, Zezyulin2013}. A novelty of \PT-symmetric systems is
that, despite gain and loss, the linear spectrum of the system can
still be all-real \cite{Bender1998,Mostafazadeh2002}, thus all
linear modes still exhibit regular wave behavior, just as in
conservative systems. This all-real linear spectrum is believed to
facilitate the existence of soliton families in \PT-symmetric
systems.

It turns out that non-\PT-symmetric dissipative systems can also
possess all-real linear spectra. Indeed, for the one-dimensional
(1D) linear Schr\"odinger operator, various non-\PT-symmetric
complex potentials with all-real spectra have been constructed by
the supersymmetry method \cite{Cooper1995,Cannata1998,Miri2013}.
Even if the linear spectrum of a non-\PT-symmetric dissipative
system is not all-real, part of that spectrum can still be real,
meaning that the system can support linear guided modes
\cite{Zezyulin2011,Siegman2003,Chen2007}. In such non-\PT-symmetric
dissipative systems, an important question is: can continuous
families of solitons bifurcate out from linear guided modes? If they
do, then the underlying dissipative physical system would allow much
more flexibility in steering nonlinear localized modes (such as
optical solitons) with continuous ranges of intensities, and this
flexibility could have potential physical applications.

In this article, we investigate the existence of soliton families in
the 1D nonlinear Schr\"odinger (NLS) equation with a
non-\PT-symmetric complex potential. This NLS system governs
paraxial nonlinear light propagation in a medium with
non-\PT-symmetric refractive-index and gain-loss landscape
\cite{Yang2010,Musslimani2008}, as well as Bose-Einstein condensates
with a non-\PT-symmetric trap and gain-loss distribution
\cite{Pitaevskii2003}. In this NLS model, we show that no soliton
families can bifurcate out from localized linear modes of a
non-periodic potential or Bloch-band edges of a periodic potential.
These results suggest that 1D non-\PT-symmetric potentials do not
support continuous families of solitons. In other words,
\PT-symmetry of a 1D complex potential is a necessary condition for
the existence of soliton families (although it is not necessary for
all-real linear spectra).

\section{Preliminaries}

The model equation we consider is the following 1D NLS equation with
a linear non-\PT-symmetric complex potential
\begin{equation}  \label{e:U}
iU_t+U_{xx}-V(x)\Psi+\sigma |\Psi|^2\Psi=0,
\end{equation}
where $V(x)$ is complex-valued and non-\PT-symmetric, i.e.,
\[ V^*(x) \ne V(-x),   \label{e:nPT} \]
the asterisk represents complex conjugation, and $\sigma=\pm 1$ is
the sign of nonlinearity. This equation governs paraxial light
transmission as well as Bose-Einstein condensates in
non-\PT-symmetric media. In this model, the nonlinearity is cubic.
But extension of our analysis to an arbitrary form of nonlinearity
is straightforward without much more effort \cite{Yang2012b}.

Regarding the non-\PT-symmetric potential $V(x)$, a remark is in
order. If this $V(x)$ is non-\PT-symmetric, but becomes
\PT-symmetric after a certain spatial translation $x_0$, i.e.,
$V(x-x_0)$ is \PT-symmetric, then wave dynamics in this
non-\PT-symmetric potential $V(x)$ is equivalent to that in the
\PT-symmetric potential $V(x-x_0)$ and is thus not the subject of
our study. Hence, in this article we require that the
non-\PT-symmetric potential $V(x)$ in Eq. (\ref{e:U}) remains
non-\PT-symmetric under any spatial translation.

For non-\PT-symmetric complex potentials, their linear spectra may
or may not contain real eigenvalues. In this article, we will
consider those potentials that admit real eigenvalues in their
linear spectra. Non-\PT potentials with all-real spectra are special
but important examples of such potentials.

We seek solitons in Eq. \eqref{e:U} of the form
\begin{equation}  \label{e:Usoliton}
U(x,t)=e^{i\mu t}u(x),
\end{equation}
where $u(x)$ is a localized function satisfying the equation
\begin{equation}  \label{e:u}
u_{xx}-V(x)u-\mu u+\sigma |u|^2u=0,
\end{equation}
and $\mu$ is a real-valued propagation constant. The question we
will investigate is, does this equation admit soliton families for a
continuous range of propagation-constant values when the potential
$V(x)$ is non-\PT-symmetric?

It is noted that Eq. (\ref{e:u}) is phase-invariant. That is, if
$u(x)$ is a solitary wave, then so is $u(x)e^{i\alpha}$, where
$\alpha$ is any real constant. In this article, solitons that are
related by this phase invariance will be considered as equivalent.

\section{Non-existence of soliton families bifurcating from
localized linear modes}  \label{sec:local}

In this section, we consider non-\PT-symmetric potentials that are
not periodic (for instance, localized potentials). Such potentials
can admit discrete real eigenvalues, i.e., linear guided modes
\cite{Cooper1995,Cannata1998,Miri2013,Zezyulin2011,Siegman2003,Chen2007}.
If this potential were real or \PT-symmetric, soliton families would
always bifurcate out from those linear guided modes. However, when
the potential is non-\PT-symmetric, we will show that such
soliton-family bifurcations are forbidden.

Suppose $V(x)$ is a non-\PT-symmetric potential which admits a
simple discrete real eigenvalue $\mu_0$, with the corresponding
localized eigenfunction $\psi(x)$, i.e.,
\[
L \psi = 0,
\]
where
\[  \label{d:L}
L \equiv \frac{d^2}{dx^2} -V(x) -\mu_0.
\]
Since $\mu_0$ is a simple eigenvalue, the equation $L\psi_g=\psi$
for the generalized eigenfunction $\psi_g$ should not admit any
solution. This means that the solvability condition of this $\psi_g$
equation should not be satisfied, i.e., its inhomogeneous term
$\psi$ should not be orthogonal to the adjoint homogeneous solution
$\psi^*$, or
\[
\langle \psi^*, \psi \rangle\ne 0,
\]
where
\[
\langle f, g \rangle \equiv \int_{-\infty}^\infty f^*(x) g(x) dx
\]
is the standard inner product.

If a soliton family in Eq. (\ref{e:u}) bifurcates out from this
localized linear eigenmode, then we can expand these solitons into a
perturbation series. We will show that this perturbation series
requires an infinite number of nontrivial conditions to be satisfied
simultaneously, which is impossible in practice due to lack of
spatial symmetries in the 1D potential $V(x)$.

To proceed, let us expand these solitons into a perturbation series
\[
u(x; \mu) = \epsilon^{1/2} \left[ u_0(x) + \epsilon u_1(x) +\epsilon^2
u_2(x) +\dots\right],   \label{e:uexpand}
\]
where $\epsilon \equiv \mu-\mu_0$ is small. Substituting this
expansion into Eq. (\ref{e:u}), at $O(\epsilon^{1/2})$ we get
\[  \label{e:u0}
Lu_0=0,
\]
hence
\[
u_0=c_0\psi,   \label{s:u0}
\]
where $c_0$ is a certain non-zero constant.

At $O(\epsilon^{3/2})$, we get the equation for $u_1$ as
\[  \label{e:u1}
Lu_1=c_0\left(\psi-\sigma |c_0|^2|\psi|^2\psi\right).
\]
Here the $u_0$ solution (\ref{s:u0}) has been utilized. The
solvability condition of this $u_1$ equation is that its right hand
side be orthogonal to the adjoint homogeneous solution $\psi^*$.
This condition yields an equation for $c_0$ as
\[  \label{f:c02}
|c_0|^2 = \frac{\langle \psi^*, \psi\rangle}{\sigma  \langle \psi^*, |\psi|^2\psi\rangle}.
\]
Here we have assumed that the denominator $\langle \psi^*,
|\psi|^2\psi\rangle \ne 0$. If it is zero, perturbation expansions
different from (\ref{e:uexpand}) would be needed, but the
qualitative result would remain the same as that given below.

Since $|c_0|$ is real and $\sigma=\pm 1$, Eq. (\ref{f:c02}) then
requires that
\[  \label{e:Q1}
Q_1 \equiv  \frac{\langle \psi^*, \psi\rangle}{\langle \psi^*, |\psi|^2\psi\rangle} \quad \mbox{must be real}.
\]
In a non-\PT-symmetric complex potential, $Q_1$ is generically
complex, thus this condition is generically not satisfied.

It turns out that Eq. (\ref{e:Q1}) is only the first condition for
soliton-family bifurcations. As we pursue the perturbation expansion
(\ref{e:uexpand}) to higher orders, infinitely more conditions will
also appear. This will be demonstrated below.

If condition (\ref{e:Q1}) is met, then the $u_1$ equation
(\ref{e:u1}) is solvable. Its solution is
\[  \label{s:u1}
u_1=\hat{u}_1+c_1\psi,
\]
where $\hat{u}_1$ is a particular solution to Eq. (\ref{e:u1}), and
$c_1$ is a constant coefficient of the homogeneous solution $\psi$.

At $O(\epsilon^{5/2})$, the $u_2$ equation is
\[  \label{e:u2}
Lu_2=u_1-\sigma(u_0^2u_1^*+2|u_0|^2u_1).
\]
Substituting the above $u_1$ solution into this equation, we get
\[
Lu_2=c_1(1-2\sigma|u_0|^2)\psi-c_1^*\sigma u_0^2\psi^*+h_2,
\]
where
\[  \nonumber
h_2\equiv (1-2\sigma|u_0|^2)\hat{u}_1-\sigma u_0^2\hat{u}_1^*.
\]
The solvability condition of this $u_2$ equation is that its right
hand side be orthogonal to the adjoint homogeneous solution
$\psi^*$. Recalling the $u_0$ solution (\ref{s:u0}) and utilizing
the solvability condition of the $u_1$ equation (\ref{e:u1}), the
solvability condition of the above $u_2$ equation then reduces to
\[
c_1+c_1^*=\frac{\langle \psi^*, h_2\rangle}{\langle \psi^*, \psi\rangle}.
\]
In order for this equation to admit $c_1$ solutions, we need to
require that
\[  \label{e:Q2}
Q_2 \equiv  \frac{\langle \psi^*, h_2\rangle}{\langle \psi^*, \psi\rangle} \quad \mbox{must be real}.
\]
This is the second condition that must be satisfied in order for the
perturbation series solution (\ref{e:uexpand}) of $u(x; \mu)$ to
exist. In a non-\PT-symmetric complex potential, this condition is
generically not satisfied either.

Carrying out this perturbative calculation to higher orders, we can
show that infinitely more conditions of the type (\ref{e:Q2}) will
appear. Due to lack of symmetry of the involved functions, it is
practically impossible for these infinite conditions to be met
simultaneously. Thus soliton families cannot bifurcate out from a
localized linear eigenmode in a non-\PT-symmetric potential.

\section{Non-existence of soliton families bifurcating from
Bloch-band edges}  \label{sec:periodic}

In this section, we consider periodic non-\PT-symmetric potentials.
According to the Bloch-Floquet theory, these potentials do not admit
discrete eigenvalues, but they possess Bloch bands which can be
partially-real or all-real \cite{Musslimani2008,Nixon2012}. In
periodic real or \PT-symmetric potentials, soliton families can
bifurcate out from edges of Bloch bands \cite{Yang2010,Nixon2012}.
However, when the periodic potential is non-\PT-symmetric, we will
show that these soliton-family bifurcations from band edges are also
forbidden.

Suppose $V(x)$ is a periodic non-\PT-symmetric complex potential
that possesses a real segment of Bloch bands, and $\mu_0$ is a
real-valued edge of this Bloch band with the corresponding Bloch
mode $p(x)$, i.e.,
\[  \label{e:p}
Lp=0,
\]
where $L$ is as defined in Eq. (\ref{d:L}). According to the
Bloch-Floquet theory, the Bloch mode $p(x)$ at edge $\mu_0$ is
either $T$- or $2T$-periodic, where $T$ is the period of the
potential $V(x)$. In addition, at the band edge, the eigenvalue
$\mu_0$ is simple, i.e., $Lp_g=p$ does not admit generalized
eigenfunctions $p_g$. This means that the inhomogeneous term $p$
should not be orthogonal to the adjoint homogeneous solution $p^*$,
i.e.,
\[
\langle p^*, p \rangle\ne 0,
\]
where the inner product here (and throughout this section) is
defined as
\[  \label{d:innprodb}
\langle f, g \rangle \equiv \int_{0}^T f^*(x) g(x) dx.
\]

Now we consider bifurcations of soliton families from this real band
edge $\mu_0$. If the potential $V(x)$ is real, this soliton-family
bifurcation has been studied in great detail in
\cite{Baizakov2002,Pelinovsky2004,Yang2010,Hwang2011}, and it was
shown that two soliton families could bifurcate out from each
Bloch-band edge. In a non-\PT-symmetric complex potential, however,
we will show below that for this soliton-family bifurcation to
occur, an infinite number of nontrivial conditions would have to be
satisfied simultaneously, which is impossible in practice.

Suppose a soliton family bifurcates out from the band edge $\mu_0$.
Then near this edge, we can expand this soliton family and its
propagation constant $\mu$ into perturbation series
\[  \label{e:uexpandb}
u(x; \mu)=\epsilon (u_0+\epsilon u_1+\epsilon^2 u_2 +\dots),
\]
\[  \label{e:muexpandb}
\mu=\mu_0+\mu_2\epsilon^2+\mu_4\epsilon^4+\dots,
\]
where $\epsilon$ is a small real parameter,
\[
u_0=A(X)p(x)
\]
is a Bloch-wave packet, $X=\epsilon x$ is the slow spatial variable
of the packet envelope $A(X)$, and $\mu_2, \mu_4, \dots$ are real
constants.

Substituting expansions (\ref{e:uexpandb})-(\ref{e:muexpandb}) into
Eq. (\ref{e:u}), the $O(\epsilon)$ equation is satisfied
automatically due to Eq. (\ref{e:p}). At $O(\epsilon^2)$, we get the
equation for $u_1$ as
\[  \label{e:u1b}
Lu_1=-2A_X p_x.
\]
The solvability condition of this $u_1$ equation is that its right
hand side be orthogonal to the adjoint homogeneous solution
$p^*(x)$, which is satisfied automatically. Thus this $u_1$ equation
is solvable. Its solution can be written as
\[
u_1=A_X \nu,
\]
where $\nu(x)$ is a periodic solution to the equation
\[
L\nu=-2p_x.
\]

At $O(\epsilon^3)$, we get the equation for $u_2$ as
\[
Lu_2=-A_{XX}\left( p+2\nu_x \right) +\mu_2Ap-\sigma |A|^2A |p|^2p.
\]
Its solvability condition is that its right hand side be orthogonal
to $p^*(x)$. This condition yields the following equation for the
envelope function $A(X)$,
\[  \label{e:A}
DA_{XX}+\mu_2A-\alpha|A|^2A=0,
\]
where
\[  \label{e:Dalpha}
D\equiv -\frac{\langle p^*, p+2\nu_x\rangle}{\langle p^*, p\rangle}, \quad
\alpha\equiv \sigma \frac{\langle p^*, |p|^2p \rangle}{\langle p^*, p\rangle}.
\]

Under the previous assumption of $V(x)$ possessing a real segment of
Bloch bands with $\mu_0$ as its edge, we can show by analyzing the
linear Bloch-wave solution of Eq. (\ref{e:u}) through perturbation
expansions near the band edge $\mu_0$ that, the constant $D$ in the
above equation (\ref{e:Dalpha}) is related to the dispersion
relation $\mu=\mu(k)$ as \cite{Pelinovsky2004,Yang2010}
\[
D=\frac{1}{2} \left. \frac{d^2\mu}{dk^2}\right|_{\mu=\mu_0},
\]
hence $D$ is real. Then in order for the envelope equation
(\ref{e:A}) to admit a localized solution, the coefficient $\alpha$
must be real. Thus, bifurcation of soliton families from a band edge
$\mu_0$ requires that
\[  \label{e:R1}
R_1 \equiv \frac{\langle p^*, |p|^2p \rangle}{\langle p^*, p\rangle} \quad \mbox{must be real}.
\]
In a non-\PT-symmetric periodic potential, $R_1$ is generically
complex, thus this condition is generically not satisfied.

Carrying this perturbation calculation to higher orders, we will
find that infinitely more non-trivial conditions also need to be
satisfied in order for soliton-family bifurcations from band edges
to occur, similar to the case of soliton bifurcations from localized
linear modes in the previous section. For instance, the next
condition, which comes from the solvability condition of the $u_3$
equation, is that
\[ \label{e:R2}
R_2 \equiv i\frac{\langle p^*, p^2\nu^*-|p|^2\nu \rangle }{\langle p^*, |p|^2p\rangle} \quad \mbox{must be real}.
\]
Due to lack of symmetry in the complex potential and its Bloch
modes, each of these infinite conditions is nontrivial and is
generically not satisfied. The requirement of them all satisfied
simultaneously is practically impossible. Thus we conclude that in a
non-\PT-symmetric periodic potential, no soliton families can
bifurcate out from Bloch-band edges either.

\section{Examples}

In this section, we corroborate the general analytical conclusions
of the previous two sections by three examples.

In these examples, non-\PT-symmetric complex potentials are obtained
by the supersymmetry method so that they have all-real spectra
\cite{Cooper1995,Cannata1998,Miri2013}. This supersymmetry method is
briefly summarized below.

\subsection{Non-\PT-symmetric potentials with
all-real spectra}

Suppose $V_1(x)$ is a potential with all-real spectrum, and
$\mu^{(1)}$ is an eigenvalue of this potential with eigenfunction
$\psi^{(1)}$, i.e.,
\[
\left[ \frac{d^2}{dx^2} -V_1(x)-\mu^{(1)}\right] \psi^{(1)}=0.
\]
We first factorize the linear operator in this equation as
\[   \label{e:factorV1}
-\frac{d^2}{dx^2} +V_1(x)+\mu^{(1)} = \left[-\frac{d}{dx}+W(x)\right]\left[\frac{d}{dx}+W(x)\right].
\]
The function $W(x)$ in this factorization can be obtained by
requiring $\psi^{(1)}$ to annihilate $d/dx+W(x)$, and this gives
$W(x)$ as
\[  \label{f:W}
W(x)=-\frac{d}{dx} \ln (\psi^{(1)}).
\]
It is easy to directly verify that this $W(x)$ does satisfy the
factorization equation (\ref{e:factorV1}).

Now we switch the two operators on the right side of the above
factorization, and this leads to a new potential $V_2(x)$,
\[  \label{e:factorV2}
-\frac{d^2}{dx^2} +V_2(x)+\mu^{(1)} = \left[\frac{d}{dx}+W(x)\right]\left[-\frac{d}{dx}+W(x)\right],
\]
where
\[  \label{f:V2}
V_2=V_1+2W_x.
\]
This $V_2$ potential is referred to as the partner potential of
$V_1$, and it has the same spectrum as $V_1$, since operators $AB$
and $BA$ share the same spectrum. The only possible exception is the
eigenvalue $\mu^{(1)}$. Indeed, using the $V_2$-factorization
(\ref{e:factorV2}) we can show that $\mu^{(1)}$ is not in the
spectrum of $V_2$ (unless $\mu^{(1)}$ is a degenerate eigenvalue of
$V_1$, i.e., its algebraic multiplicity is higher than its geometric
multiplicity in the $V_1$ potential).

The new potential $V_2$, however, is only real or \PT-symmetric if
$V_1$ is so. In order to derive non-\PT-symmetric potentials, we
build a new factorization for the $V_2$ potential,
\[  \label{e:factorV2new}
-\frac{d^2}{dx^2}+V_2(x)+\mu^{(1)} =
\left[\frac{d}{dx}+\widetilde{W}(x)\right]\left[-\frac{d}{dx}+\widetilde{W}(x)\right].
\]
Using the previous $V_2$ factorization (\ref{e:factorV2}), the
function $\widetilde{W}$ in this new factorization can be derived as
\cite{Miri2013}
\[  \label{f:Wtilde}
\widetilde{W}(x)=-\frac{d}{dx} \ln (\widetilde{\psi}^{(1)}),
\]
where
\[  \label{f:psi1tilde}
\widetilde{\psi}^{(1)}(x) = \frac{\psi^{(1)}(x)}{c+\int_0^x [\psi^{(1)}(\xi)]^2 d\xi},
\]
and $c$ is an arbitrary complex constant. For this new $V_2$
factorization, its partner potential, defined through
\[  \label{e:factorV1tilde}
-\frac{d^2}{dx^2} +\widetilde{V}_1(x)+\mu^{(1)} = \left[-\frac{d}{dx}+\widetilde{W}(x)\right]\left[\frac{d}{dx}+\widetilde{W}(x)\right],
\]
is then
\[
\widetilde{V}_1=V_2-2\widetilde{W}_x.
\]
Utilizing the $V_2$ and $\widetilde{W}$ formulae (\ref{f:V2}) and
(\ref{f:Wtilde}), this $\widetilde{V}_1$ potential is then found to
be
\[  \label{f:V1tilde}
\widetilde{V}_1(x)=V_1(x)-2\frac{d^2}{dx^2}\ln\left[c+\int_0^x
[\psi^{(1)}(\xi)]^2 d\xi\right].
\]
For generic values of the complex constant $c$, this
$\widetilde{V}_1$ potential is complex and non-\PT-symmetric. In
addition, its spectrum is identical to that of $V_1$. Indeed, even
though $\mu^{(1)}$ may not lie in the spectrum of $V_2$, it is in
the spectrum of $\widetilde{V}_1$ with eigenfunction
$\widetilde{\psi}^{(1)}$. Hence if $V_1$ has an all-real spectrum,
so does $\widetilde{V}_1$. Notice that this $\widetilde{V}_1$
potential, referred to as superpotential below, is actually a family
of potentials due to the free complex constant $c$.

If the original potential $V_1$ is localized, taking $\psi^{(1)}$ as
any of its discrete eigenmodes would lead to a localized
superpotential. However, if we want to construct a periodic
superpotential from a periodic original potential $V_1$, then it is
easy to see from Eq. (\ref{f:V1tilde}) and the Bloch-Floquet theory
that the Bloch mode $\psi^{(1)}$ must be $T$- or $2T$-periodic, and
\[  \label{c:psi2zero}
\int_0^T [\psi^{(1)}(x)]^2 dx=0,
\]
where $T$ is the period of the $V_1$ potential. The former
requirement means that the Bloch mode $\psi^{(1)}$ is located at the
center or edge of the Brillouin zone. The latter requirement
(\ref{c:psi2zero}) means that the Bloch mode $\psi^{(1)}$ must be
complex, hence so is the $V_1$ potential. In addition, $\langle
\psi^{(1)*}, \psi^{(1)}\rangle=0$, thus this Bloch mode is
degenerate. At such a degenerate point, the local dispersion curve
is two lines crossing each other like `$\times$'. Due to this
degeneracy, when the $V_1$ potential is perturbed, complex
eigenvalues will bifurcate out from $\mu^{(1)}$ \cite{Nixon2012}.
Thus if the $V_1$ potential is \PT-symmetric, then it must be at the
phase-transition point (also known as \PT-symmetry-breaking point)
\cite{Bender1998,Musslimani2008,Nixon2012}.

\subsection{Three examples}
Now we consider three examples of non-\PT-symmetric superpotentials
with all-real spectra, and show that the conditions for
soliton-family bifurcations in them are not satisfied. Of these
three examples, the first two pertain to localized superpotentials,
and the third to periodic superpotentials.

\vspace{0.1cm} \textbf{Example 1}\ \ In our first example, the
superpotential (\ref{f:V1tilde}) is created from the harmonic
potential
\[  \label{f:V1example1}
V_1(x)=x^2
\]
and its first eigenmode of $\mu^{(1)}=-1$ with
\[
\psi^{(1)}=e^{-x^2/2}.
\]
In other words, the superpotential (\ref{f:V1tilde}) is
\[  \label{f:Vexample1}
V(x)=x^2-2\frac{d^2}{dx^2}\ln\left[c+\int_0^x e^{-\xi^2}d\xi\right],
\]
where $c$ is a complex constant.
When $c$ is real, so is $V(x)$. When $c$ is purely imaginary, $V(x)$
is complex and \PT-symmetric. For all other $c$ values, the
superpotential (\ref{f:Vexample1}) is complex and non-\PT-symmetric.
An example of this non-\PT-symmetric superpotential with $c=1+i$ is
illustrated in Fig. 1(a). The spectrum of this superpotential (for
all $c$ values) is $\{-1, -3, -5, \dots\}$, which is the same as
that of the harmonic potential (\ref{f:V1example1}).

For this superpotential (\ref{f:Vexample1}), we consider the
bifurcation of soliton families from its first eigenmode of
$\mu_0=-1$, whose eigenfunction is that given in Eq.
(\ref{f:psi1tilde}), i.e.,
\[
\psi=\frac{e^{-x^2/2}}{c+\int_0^x e^{-\xi^2}d\xi}.
\]
Substituting this eigenmode into the $Q_1$ condition (\ref{e:Q1}),
we find that this condition is never satisfied for any complex $c$
value that is not real or purely imaginary. For instance, if the
imaginary part of $c$ is fixed as one, then the imaginary part of
$Q_1$ versus the real part of $c$ is plotted in Fig. 1(b). One can
see that $\mbox{Im}(Q_1)\ne 0$ when $\mbox{Re}(c)\ne 0$, indicating
that $Q_1$ is never real when the superpotential (\ref{f:Vexample1})
is non-\PT-symmetric; thus condition (\ref{e:Q1}) is not satisfied.
As a consequence, bifurcation of soliton families from the first
eigenmode of the non-\PT-symmetric superpotential
(\ref{f:Vexample1}) cannot take place.

\begin{figure}[tb!]
\includegraphics[width=0.5\textwidth]{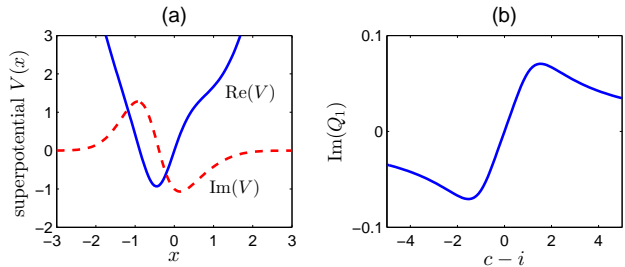}

\smallskip
\caption{(Color online) (a) Superpotential (\ref{f:Vexample1}) with $c=1+i$; (b) imaginary part of $Q_1$ in Eq. (\ref{e:Q1})
for various complex values of $c$ with $\mbox{Im}(c)=1$. }
\end{figure}

\vspace{0.1cm} \textbf{Example 2}\ \ In our second example, the
superpotential (\ref{f:V1tilde}) is created from a \PT-symmetric
double-well potential
\begin{eqnarray}  \label{f:V1example2}
V_1(x)=-3[\mbox{sech}^2(x+1)+\mbox{sech}^2(x-1)]  \hspace{1cm} \nonumber \\
 +0.5i[\mbox{sech}^2(x+1)-\mbox{sech}^2(x-1)]. \hspace{0.5cm}
\end{eqnarray}
This $V_1$ potential has an all-real spectrum that contains three
positive discrete eigenvalues and a continuous spectrum of
$(-\infty, 0]$. Its first discrete eigenvalue is $\mu^{(1)}\approx
2.3687$, and the eigenfunction $\psi^{(1)}$ of this first eigenvalue
will be used to build the superpotential (\ref{f:V1tilde}).

This superpotential is always complex, and is non-\PT-symmetric if
$c$ is not purely imaginary. When $c=4-i$, this superpotential is
illustrated in Fig. 2(a).

For this superpotential (with arbitrary $c$), we also consider the
bifurcation of soliton families from its first eigenmode
$\widetilde{\psi}^{(1)}$, whose eigenvalue $\mu^{(1)}$ is as given
above. In the notations of our analysis in Sec. \ref{sec:local}, we
choose
\[ \label{e:mu0psiexample2}
\mu_0=\mu^{(1)}, \quad \psi=\widetilde{\psi}^{(1)}.
\]
Here the formula for $\widetilde{\psi}^{(1)}$ is provided by Eq.
(\ref{f:psi1tilde}), where $\psi^{(1)}$ is the first eigenmode of
the original double-well potential $V_1$, which can be obtained
numerically.

Substituting eigenmode $\psi$ of (\ref{e:mu0psiexample2}) into the
$Q_1$ formula (\ref{e:Q1}), we find that in the complex $c$-plane,
this $Q_1$ is non-real everywhere except on the imaginary axis and
on a certain quasi-ellipse. The $c$ values on the imaginary axis
only yield \PT-symmetric superpotentials and are not our concern.
For $c$ values on that quasi-ellipse, the superpotential is
non-\PT-symmetric and $Q_1$ is real, thus the first condition
(\ref{e:Q1}) for soliton-family bifurcations is satisfied. However,
we have found that on that $c$-ellipse, the second condition
(\ref{e:Q2}) is not met, thus this soliton-family bifurcation cannot
occur.

To illustrate, we fix $\mbox{Im}(c)=-1$. Then $\mbox{Im}(Q_1)$
versus $\mbox{Re}(c)$ is plotted in Fig. 2(b). For non-\PT-symmetric
superpotentials, $\mbox{Re}(c)\ne 0$. Then we see that at
$\mbox{Re}(c)\approx \pm 1.2918$ (marked by red dots in that
figure), $\mbox{Im}(Q_1)=0$, i.e., $Q_1$ is real. These two $c$
values, $\pm 1.2918-i$, are on that $c$-ellipse mentioned above. But
at these two $c$ values, we have checked that $Q_2$ is not real,
thus the second condition (\ref{e:Q2}) for soliton-family
bifurcations is not met.

\begin{figure}[tb!]
\includegraphics[width=0.5\textwidth]{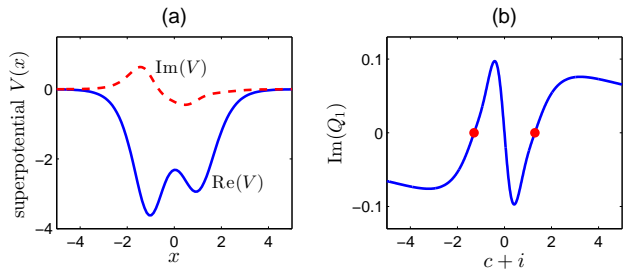}

\smallskip
\caption{(Color online) (a) Superpotential (\ref{f:V1tilde}) built from a double-well potential (\ref{f:V1example2}) and its first discrete eigenmode
with $c=4-i$; (b) imaginary part of $Q_1$ in Eq. (\ref{e:Q1}) for various
complex values of $c$ with $\mbox{Im}(c)=-1$. }
\end{figure}

\vspace{0.1cm} \textbf{Example 3}\ \ Our third example pertains to a
periodic superpotential. In view of the discussions in the end of
the previous subsection, this periodic superpotential
(\ref{f:V1tilde}) can be built from the original \PT-symmetric
periodic potential
\[ \label{f:V1example3}
V_1(x)=-V_0^2 e^{2ix},
\]
and its Bloch mode
\[
\psi^{(1)}=I_1(V_0e^{ix})
\]
with eigenvalue $\mu^{(1)}=-1$. Here $V_0$ is a real constant, and
$I_n$ is the modified Bessel function. It is known that this $V_1$
potential is at the phase transition point
\cite{Musslimani2008,Nixon2012}, and its Bloch mode $\psi^{(1)}$ is
located at the edge of the first Bloch band with a `$\times$'-shaped
local dispersion curve \cite{Bender1999,Nixon2012b}. The resulting
periodic superpotential (\ref{f:V1tilde}) is
\[  \label{f:Vexample3}
V(x)=-V_0^2 e^{2ix}-2\frac{d^2}{dx^2}\ln\left[c+\int_0^x  I_1^2(V_0 e^{i\xi}) d\xi\right],
\]
where $c$ is a complex constant.

This superpotential (\ref{f:Vexample3}) is $\pi$-periodic, and is
non-\PT-symmetric as long as $c$ is not purely imaginary. When
$c=0.5-2i$ and $V_0=1$, this superpotential is illustrated in
Fig.~3(a).

The dispersion relation of this superpotential (for all $c$ values)
is the same as that of the original potential (\ref{f:V1example3}),
i.e.,
\[
\mu=-(k+2m)^2,
\]
where $k$ is in the Brillouin zone $[-1, 1]$, and $m$ is any
non-negative integer \cite{Nixon2012b}. From this dispersion
relation, we see that Bloch bands of this superpotential cover the
entire interval of $-\infty < \mu < 0$. Thus the only possible band
edge for soliton bifurcations is $\mu_0=0$ (upper edge of the first
Bloch band with $k=0$). At this band edge, the Bloch mode in the
original $V_1$ potential is
\[
p^{(1)}(x)=I_0(V_0e^{ix}).
\]
Then the corresponding Bloch mode in the superpotential
(\ref{f:Vexample3}) can be derived from Eqs. (\ref{e:factorV1}),
(\ref{e:factorV2}), (\ref{e:factorV2new}) and
(\ref{e:factorV1tilde}) as \cite{Miri2013}
\[
p=\left(-\frac{d}{dx} + \widetilde{W}\right) \left( \frac{d}{dx} +W\right) p^{(1)},
\]
where $W$ and $\widetilde{W}$ are given by Eqs. (\ref{f:W}) and
(\ref{f:Wtilde}).

Substituting this Bloch mode $p(x)$ into the $R_1$ formula
(\ref{e:R1}), we find that this $R_1$ is non-real everywhere in the
complex $c$-plane, except for the imaginary axis and a certain
quasi-circle. The $c$ values on the imaginary axis lead to
\PT-symmetric superpotentials which are irrelevant for our study.
For $c$ values on that quasi-circle, $R_1$ is real, but $R_2$ in Eq.
(\ref{e:R2}) is non-real, thus the second condition (\ref{e:R2}) is
not met. As a consequence, soliton-family bifurcations from this
Bloch-band edge $\mu_0=0$ cannot occur. This situation is similar to
that in Example 2.

For demonstration purpose, we fix $\mbox{Im}(c)=-2$ and $V_0=1$.
Then $\mbox{Im}(R_1)$ versus $\mbox{Re}(c)$ is plotted in Fig. 3(b).
We see that on this line of $c$ values, $\mbox{Im}(R_1)\ne 0$ when
$\mbox{Re}(c)\ne 0$, indicating that $R_1$ is non-real when the
superpotential (\ref{f:Vexample3}) is non-\PT-symmetric, hence the
first condition (\ref{e:R1}) for soliton-family bifurcations is not
met.

\begin{figure}[tb!]
\includegraphics[width=0.5\textwidth]{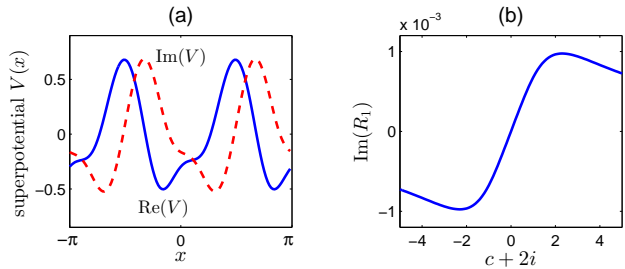}

\smallskip
\caption{(Color online) (a) Periodic superpotential (\ref{f:Vexample3}) with $c=0.5-2i$ and $V_0=1$; (b) imaginary part of $R_1$ in Eq. (\ref{e:R1})
for various complex values of $c$ with $\mbox{Im}(c)=-2$ and $V_0=1$. }
\end{figure}

\section{Summary and discussion}

In this article, we have shown that for the 1D NLS equation with a
non-\PT-symmetric periodic or non-periodic potential, no continuous
families of solitons can bifurcate out from linear modes of the
potential, even if this potential has an all-real spectrum. This
analytical finding is also corroborated by several specific examples
containing complex superpotentials with all-real spectra. This
result suggests that \PT-symmetry of a 1D complex potential is a
necessary condition for the existence of soliton families. This
conclusion highlights the importance of \PT-symmetry for the study
of nonlinear soliton states, even though it is not necessary for
all-real linear spectrum.

If a complex potential is \PT-symmetric, then repeating the
perturbative calculations in sections \ref{sec:local} and
\ref{sec:periodic} of this article, we will find that those infinite
conditions, such as (\ref{e:Q1}), (\ref{e:Q2}), (\ref{e:R1}) and
(\ref{e:R2}), are all automatically satisfied due to \PT-symmetry of
the potential and other involved functions. For instance, for
\PT-symmetric non-periodic potentials, the linear eigenmode $\psi$
and solutions $u_0$, $\hat{u}_1$ in Sec. \ref{sec:local} can be made
\PT-symmetric through phase invariance. Thus quantities $Q_1$, $Q_2$
in Eqs. (\ref{e:Q1}), (\ref{e:Q2}) are automatically real, making
conditions (\ref{e:Q1}) and (\ref{e:Q2}) automatically fulfilled. As
a consequence, soliton families \emph{can} be successfully
constructed from perturbation expansions. This analytical result is
consistent with earlier numerical reports of soliton families in
various \PT-symmetric potentials \cite{Wang2011, Lu2011, Nixon2012,
Zezyulin2012a}. Combining this result with the finding of this
article, we argue that in the 1D NLS equation with a complex
potential, \PT-symmetry of the potential is both necessary and
sufficient for the existence of soliton families (assuming that this
potential admits real discrete eigenvalues or real Bloch bands).
Soliton families that exist in a 1D \PT-symmetric potential are
always \PT-symmetric, as was shown recently in \cite{Yang2013}.

\vspace{0.2cm} \textbf{Acknowledgment:} The author thanks Prof. V.V.
Konotop for helpful discussions. This work was supported in part by
the Air Force Office of Scientific Research (Grant USAF
9550-12-1-0244) and the National Science Foundation (Grant
DMS-1311730).


\end{document}